\documentclass[prc,twocolumn,showpacs,amsmath,amssymb]{revtex4}

\usepackage{times}
\usepackage{graphicx}
\usepackage{dcolumn}
\usepackage{bm}
\usepackage{amstext}

\begin{document}

\title{Superdeformed and hyperdeformed states in Z=122 isotopes}
\author{S. K. Patra$^1$, M. Bhuyan$^{1,2}$, M. S. Mehta$^3$ and Raj K. Gupta$^3$}
\affiliation{$^1$ Institute of Physics, Sachivalaya Marg, Bhubaneswar-751 005, India. \\ 
$^2$ School of Physics, Sambalpur University, Jyotivihar, Burla-768 019, India.\\
$^3$ Department of Physics, Panjab University, Chandigarh-160 014, India.
}

\date{\today}

\begin{abstract}
We calculate the binding energy, root-mean-square radius and quadrupole deformation parameter for the recent, possibly 
discovered superheavey element Z=122, using the axially deformed relativistic mean field (RMF) and non-relativistic Skyrme 
Hartree-Fock (SHF) formalisms. The calculation is extended to include various isotopes of Z=122 element, strarting from 
A=282 to A=320. We predict highly deformed structures in the ground state for all the isotopes. A shape transition appears 
at about A=290 from a highly oblate to a large prolate shape, which may be considered as the superdeformed and hyperdeformed 
structures of Z=122 nucleus in the mean field approaches. The most stable isotope (largest binding energy per nucleon) is 
found to be $^{302}$122, instead of the experimentally observed $^{292}$122.
\end{abstract}


\maketitle

\section{Introduction}

The stability of nuclei in superheavy mass region was predicted in mid sixties \cite{myers65,sobi66,mosel69} when shell 
correction was added to the liquid drop binding energy and the possible shell closure was pointed out at Z=114 and N=184. 
Myers and Swiatecki \cite{myers66} concluded that the half-lives of nuclei near the shell closures must be long enough to 
get observed. In other words, nuclei with zero shell effects would not be stable and decay immediately, as was predicted by
macroscopic liquid drop models for Z$>$100 nuclides. Recentally, however, the spectroscopic studies of the nuclei beyond 
Z=100 have become possible \cite{herz04}, and the heaviest nucleus studied so far in this series of experiments 
\cite{herz06} is $^{254}$No (Z=102, N=152). Thus, the progress in experimental techniques has drawn our attention and 
opened up the field once again for further theoretical investigations in structure physics of nuclei in the superheavy 
mass region.

Even though, experimentally, the elements upto Z=118 have been synthesized to-date, with half-lives varying from few minutes 
to milliseconds \cite{hof00,oga07}, the above mentioned theoretically predicted center of island of stability could not be
located precisely. Recently, more microscopic theoretical calculations have predicted various other regions of stability, 
namely, Z=120, N=172 or 184 \cite{rutz97,gupta97,patra1} and Z=124 or 126, N=184 \cite{cwiok96,kruppa00}. Apparently, there 
is a need to design the new experiments to solve the outstanding problem of locating the precise island of stability for 
superheavy elements. In an effort in this direction, using inductively coupled plasma-sector field mass spectroscopy,
Marinov {\it et al.} \cite{marinov07} have observed some neutron-deficient Th isotopes in naturally occuring Thorium 
substances. Long-lived isomeric states, with estimated half-lives $T_{1/2}\ge$10$^8$ y, have been identified in the 
neutron-deficient $^{211,213,217,218}$Th isotopes, which are associated with the superdeformed (SD) or hyperdeformed (HD) 
states (minimma) in potential energy surfaces (PES). In a similar search for long-lived, trans-actinides in natural 
materials, more recently, these authors \cite{marinov09} obtained a possible evidence for the existence of a long-lived 
superheavy nucleus  with mass number A=292 and atomic number Z=122 in natural Thorium. The half life is again estimated to 
be the same as above, i.e. $T_{1/2}\ge$10$^8$ y and abundance (1-10)$\times$10$^{-12}$ relative to $^{232}$Th. This 
possibility of an extremely heavey Z nucleus motivated us to see the structures of such nuclei in an isotopic mass chain. 
Therefore, based on the relativistic mean-field (RMF) and non-relativistic Skyrme Hartree-Fock (SHF) methods, we calculated 
the bulk proporties of Z=122 nucleus in an isotopic chain of mass A=282-320. This choice of mass range covers both the
predicted neutron magic numbers N=172 and 184.

The paper is organised as follows: Section II gives a brief description of the relativistic and non-relativistic mean-field 
formalisms. The effects of pairing for open shell nuclei, included in our calculations, are also discussed in this section. 
The results of our calculations are presented in Section III, and a summary of the results obtained, together with 
concluding remarks, are given in the last Section IV. 

\section{Theoretical framework}

\subsection{The Skyrme Hartree-Fock (SHF) method}

The general form of the Skyrme effective interaction, used in the mean-field models, can be expressed as an energy density 
functional $\cal H$, given as a function of some empirical parameters \cite{cha97,stone07}, as 
\begin{equation}
{\mathcal H}={\mathcal K}+{\mathcal H}_0+{\mathcal H}_3+ {\mathcal H}_{eff}+\cdots 
\label{eq:1}
\end{equation}
where ${\cal K}$ is the kinetic energy term, ${\cal H}_0$ the zero range, ${\cal H}_3$ the density dependent and
${\cal H}_{eff}$ the effective-mass dependent terms, which are relevant for calculating the properties of nuclear matter.
These are functions of 9 parameters $t_i$, $x_i$ ($i=0,1,2,3$) and $\eta$, given as
\begin{eqnarray}
{\mathcal H}_0&=&\frac{1}{4}t_0\left[(2+x_0)\rho^2 - (2x_0+1)(\rho_p^2+\rho_n^2)\right],
\label{eq:2}
\\
{\mathcal H}_3&=&\frac{1}{24}t_3\rho^\eta \left[(2+x_3)\rho^2 - (2x_3+1)(\rho_p^2+\rho_n^2)\right],
\label{eq:3}
\\
{\mathcal H}_{eff}&=&\frac{1}{8}\left[t_1(2+x_1)+t_2(2+x_2)\right]\tau \rho \nonumber \\
                  &&+\frac{1}{8}\left[t_2(2x_2+1)-t_1(2x_1+1)\right](\tau_p \rho_p+\tau_n \rho_n). \nonumber \\
\label{eq:4}
\end{eqnarray}
The kinetic energy ${\cal K}=\frac{\hbar^2}{2m}\tau$, a form used in the Fermi gas model for non-interacting fermions. Here, 
$m$ is the nucleon mass. The other terms, representing the surface contributions of a finite nucleus with $b_4$ and 
$b^{\prime}_4$ as additional parameters, are
\begin{eqnarray}
{\mathcal H}_{S\rho}&=&\frac{1}{16}\left[3t_1(1+\frac{1}{2}x_1)-t_2(1+\frac{1}{2}x_2)\right](\vec{\nabla}\rho)^2 \nonumber\\
                  &&-\frac{1}{16}\left[3t_1(x_1+\frac{1}{2})+t_2(x_2+\frac{1}{2})\right] \nonumber\\
                  &&\times\left[(\vec{\nabla}\rho_n)^2+(\vec{\nabla}\rho_p)^2\right],
                  \text{ and}
\label{eq:5}
\\
{\mathcal H}_{S\vec{J}}&=&-\frac{1}{2}\left[{b_4}\rho\vec{\nabla}\cdot\vec{J}+{b^{\prime}_4}(\rho_n\vec{\nabla}\cdot\vec{J_n}
+\rho_p\vec{\nabla}\cdot\vec{J_p})\right].
\label{eq:6}
\end{eqnarray}
Here, the total nucleon number density $\rho=\rho_n+\rho_p$, the kinetic energy density $\tau=\tau_n+\tau_p$, and the
spin-orbit density $\vec{J}=\vec{J}_n+\vec{J}_p$. The subscripts $n$ and $p$ refer to neutron and proton, respectively. 
The $\vec{J}_q=0$, $q=n$ or $p$, for spin-saturated nuclei, i.e., for nuclei with major oscillator shells completely filled. 
The total binding energy (BE) of a nucleus is the integral of the energy density functional $\cal H$.

At least eighty-seven parametrizations of the Skyrme interaction are published since 1972 (see, e.g., \cite{stone03}). 
In most of the Skyrme parameter sets, the coefficients of the spin-orbit potential $b_4=b^{\prime}_4=W_0$ \cite{rei92}, but 
we have used here the Skyrme SkI4 set with $b_4\ne b^{\prime}_4$ \cite{rei95}. This parameter set is designed for 
considerations of proper spin-orbit interaction in finite nuclei, related to the isotope shifts in Pb region. 

\subsection{The relativistic mean-field (RMF) method}

The relativistic Lagrangian density for a nucleon-meson many-body system \cite{sero86,ring90},
%%\begin{widetext}
\begin{eqnarray}
{\cal L}&=&\overline{\psi_{i}}\{i\gamma^{\mu}
\partial_{\mu}-M\}\psi_{i}
+{\frac12}\partial^{\mu}\sigma\partial_{\mu}\sigma
-{\frac12}m_{\sigma}^{2}\sigma^{2}\nonumber\\
&& -{\frac13}g_{2}\sigma^{3} -{\frac14}g_{3}\sigma^{4}
-g_{s}\overline{\psi_{i}}\psi_{i}\sigma-{\frac14}\Omega^{\mu\nu}
\Omega_{\mu\nu}\nonumber\\
&&+{\frac12}m_{w}^{2}V^{\mu}V_{\mu}
+{\frac14}c_{3}(V_{\mu}V^{\mu})^{2} -g_{w}\overline\psi_{i}
\gamma^{\mu}\psi_{i}
V_{\mu}\nonumber\\
&&-{\frac14}\vec{B}^{\mu\nu}.\vec{B}_{\mu\nu}+{\frac12}m_{\rho}^{2}{\vec
R^{\mu}} .{\vec{R}_{\mu}}
-g_{\rho}\overline\psi_{i}\gamma^{\mu}\vec{\tau}\psi_{i}.\vec
{R^{\mu}}\nonumber\\
&&-{\frac14}F^{\mu\nu}F_{\mu\nu}-e\overline\psi_{i}
\gamma^{\mu}\frac{\left(1-\tau_{3i}\right)}{2}\psi_{i}A_{\mu} .
\end{eqnarray}
%%\end{widetext}
All the quantities have their usual well known meanings. From the above Lagrangian we obtain the field equations for
the nucleons and mesons. These equations are solved by expanding the upper and lower components of the Dirac spinors and
the boson fields in an axially deformed harmonic oscillator basis with an initial deformation $\beta_{0}$. The set of 
coupled equations is solved numerically by a self-consistent iteration method. The centre-of-mass motion energy correction 
is estimated by the usual harmonic oscillator formula $E_{c.m.}=\frac{3}{4}(41A^{-1/3})$. The quadrupole deformation 
parameter $\beta_2$ is evaluated from the resulting proton and neutron quadrupole moments, as 
$Q=Q_n+Q_p=\sqrt{\frac{16\pi}5} (\frac3{4\pi} AR^2\beta_2)$. The root mean square (rms) matter radius is defined as 
$\langle r_m^2\rangle={1\over{A}}\int\rho(r_{\perp},z) r^2d\tau$, where $A$ is the mass number, and $\rho(r_{\perp},z)$ 
is the deformed density. The total binding energy and other observables are also obtained by using the standard 
relations, given in \cite{ring90}. We use the well known NL3 parameter set \cite{lala97}. This set not only 
reproduces the properties of stable nuclei but also well predicts for those far from the $\beta$-stability valley. 
As outputs, we obtain different potentials, densities, single-particle energy levels, radii, deformations and the binding
energies. For a given nucleus, the maximum binding energy corresponds to the ground state and other solutions are obtained
as various excited intrinsic states. 

\subsection{Pairing Effect }
Pairing is a crucial quantity for open shell nuclei in determining the nuclear properties. The constant gap, BCS-pairing 
approach is reasonably valid for nuclei in the valley of $\beta$-stability line. However, this approach breaks down when the 
coupling of the continum becomes important. In the present study, we deal with nuclei on the valley of stability line since 
the superheavy elements, though very exotic in nature, lie on the $\beta$-stability line. These nuclei are unstable, because 
of the repulsive Coulomb force, but the attractive nuclear shell effects come to their resque, making the superheavy element
possible to be synthesized, particularly when a combination of magic proton and neutron number happens to occur (largest 
shell correction). In order to take care of the pairing effects in these nuclei, we use the constant gap for proton and 
neutron, as given in \cite{madland81}:  $\triangle_p =RB_s e^{sI-tI^2}/Z^{1/3}$ and 
$\triangle_n =RB_s e^{-sI-tI^2}/A^{1/3}$, with $R$=5.72, $s$=0.118, $t$= 8.12, $B_s$=1, and $I = (N-Z)/(N+Z)$. This type of 
prescription for pairing effects, both in RMF and SHF, has already been used by us and many others authors \cite{patra01}. 
For this pairing approach, it is shown \cite{patra01,lala99} that the results for binding energies and quadrople 
deformations are almost identical with the predictions of relativistic Hartree-Bogoliubov (RHB) approach.

\section{Results and Discussion}

{\it Ground state properties using the SHF and RMF models:}\\
There exists a number of parameter sets for solving the standard SHF Hamiltonians and RMF Lagrangians. In many of our 
previous works and of other authors \cite{patra1,ring90,lala97,patra2,patra3,patra4} the ground state properties, like the 
binding energies (BE), quadrupole deformation parameters $\beta_2$, charge radii ($r_c$), and other bulk properties, are 
evaluated by using the various non-relativistic and relativistic parameter sets. It is found that, more or less, most of 
the recent parameter sets reproduce well the ground state properties, not only of stable normal nuclei but also of exotic 
nuclei which are far away from the valley of $\beta$-stability. This means that if one uses a reasonably acceptable 
parameter set, the predictions of the model will remain nearly force independent. 

\begin{figure}[ht]
\vspace{1.7cm}
\begin{center}
\includegraphics[width=1.0\columnwidth]{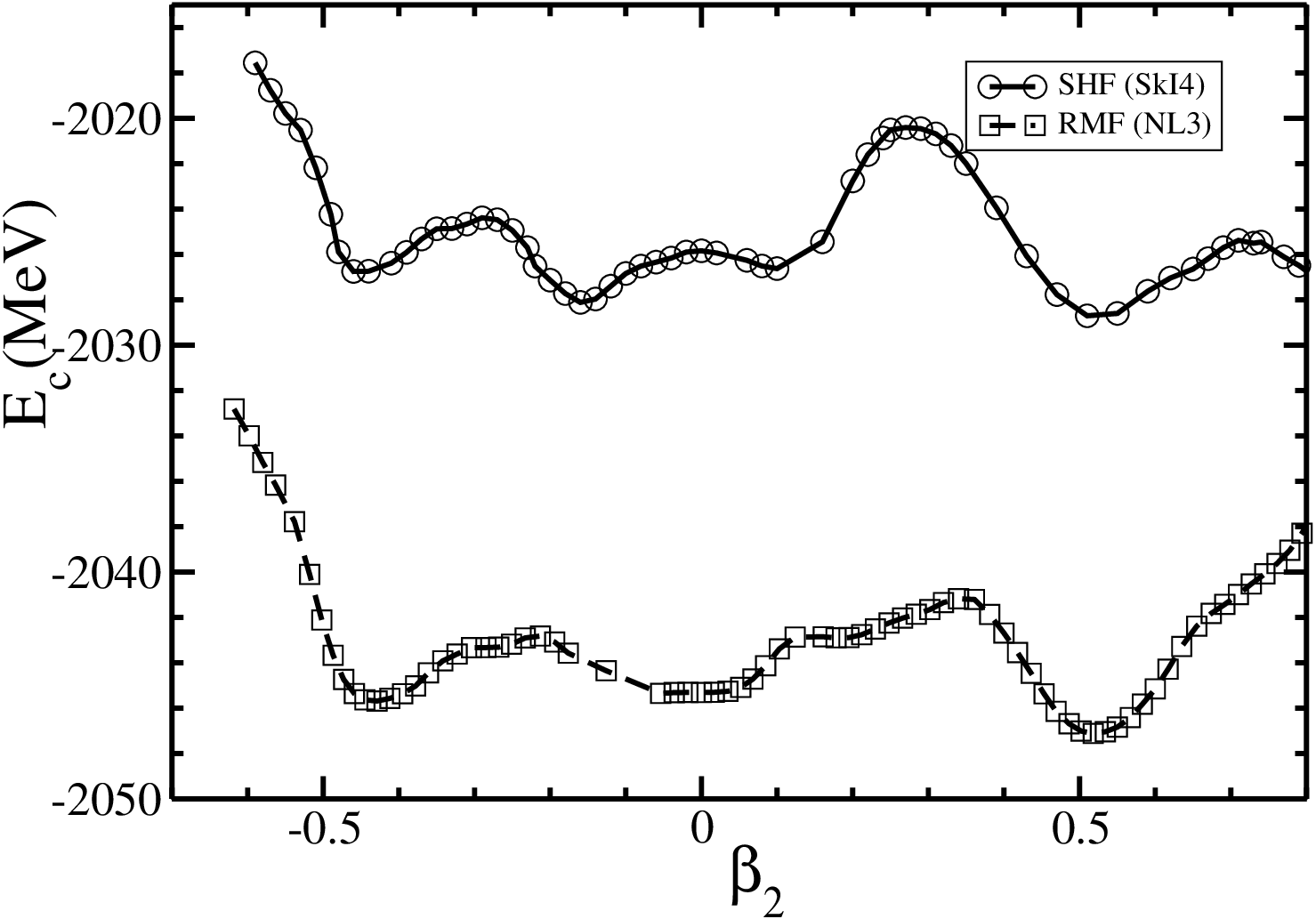}
\caption{The potential energy surfaces for $^{292}$122 nucleus as a function of quadrupole deformation parameter. The  
circles with solid-line is for SHF using SkI4 parameter set, and the squares with dash-line is for RMF calculations using
NL3 parameter set.
}
\end{center}
\label{Fig. 1}
\end{figure}

\subsection {Potential energy surface}

In this subsection, we first calculate the potential energy surfaces (PES) by using both the RMF and SHF theories in a 
constrained calculation \cite{patra4,flocard73,koepf88,reinhard89,hirata88}, i.e., instead of minimizing the $H_0$, we have 
minimized $H'=H_0-\lambda Q_{2}$, with $\lambda$ as a Lagrange multiplier and $Q_2$, the quadrupole moment. Thus, we 
calculate the binding energy corresponding to the solution at a given quadrupole deformation. Here, $H_0$ is the Dirac mean 
field Hamiltonian (the notations are standard and its form can be seen in Refs. \cite{ring90,koepf88,hirata88}) for RMF 
model and it is a Schr\"odinger mean field Hamiltonian for SHF model. In other words, we get the constrained binding energy 
from $E_c=\sum_{ij}\frac{<\psi_i|H_0-\lambda Q_2|\psi_j>}{<\psi_i|\psi_j>}$ and the ``free energy" from  
$BE=\sum_{ij}\frac{<\psi_i|H_0|\psi_j>}{<\psi_i|\psi_j>}$. In our calculations, the free energy solution does not depend on 
the initial guess value of the basis deformation $\beta_0$ as long as it is nearer to the minimum in PES. However it 
converges to some other local minimum when $\beta_0$ is drastically different, and in this way we evaluate a different 
isomeric state for a given nucleus.

\begin{figure}[ht]
\vspace{1.5cm}
\begin{center}
\includegraphics[width=1.0\columnwidth]{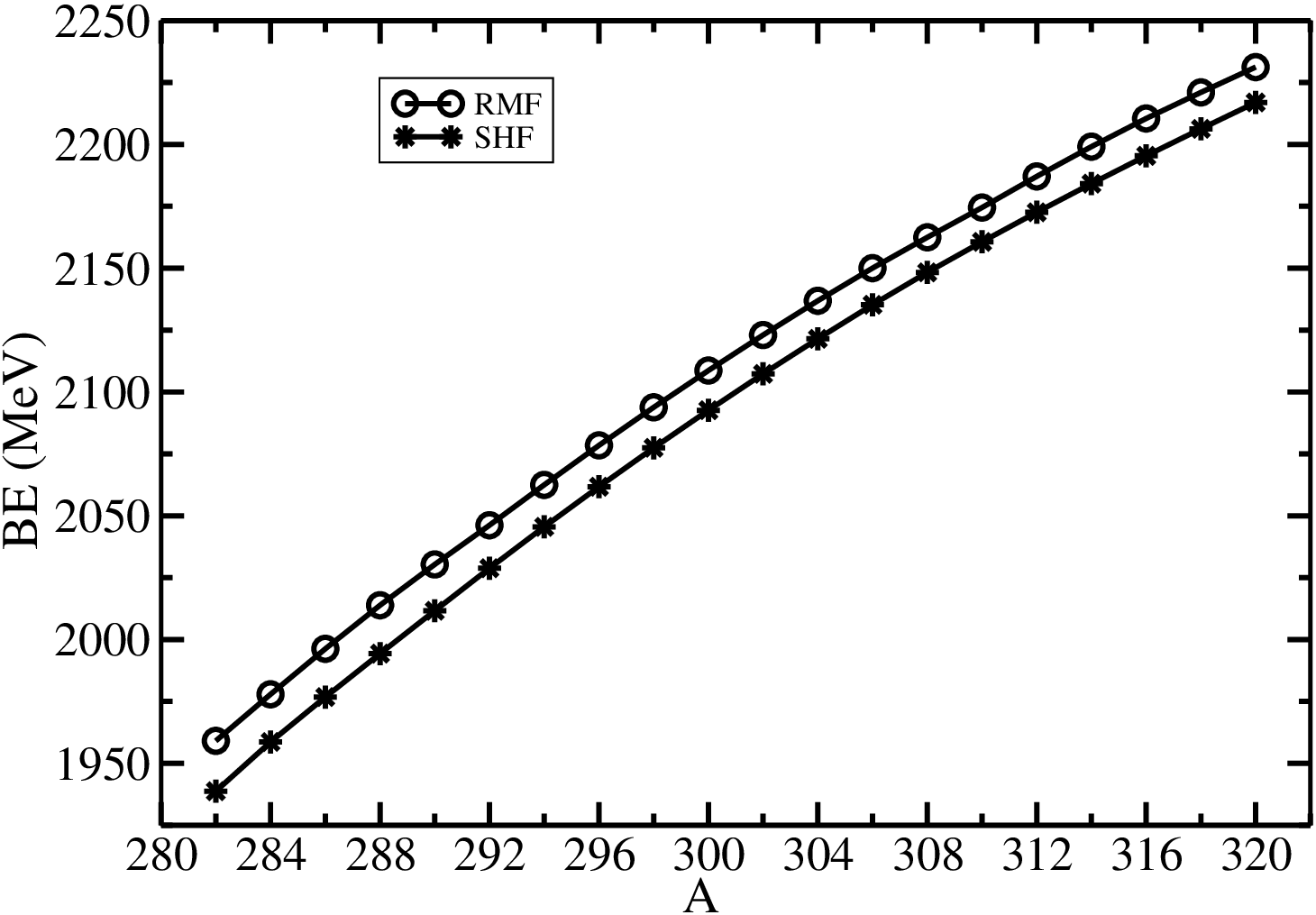}
\caption{The total binding energy for $^{282-320}$122 nuclei in SHF(SkI4) and RMF(NL3) calculations.
}
\end{center}
\label{Fig. 2}
\end{figure}

The PES, i.e., the potential energy as a function of quadrupole deformation parameter $\beta_2$, for the superheavy nucleus 
$^{292}$122, is shown in Fig. 1. Both the RMF and SHF results are given for comparisons. The calculated PES is shown for a 
wide range of oblate to prolate deformations. We notice from this figure that in RMF, minima appear at around $\beta_2$= 
-0.436, -0.032 and 0.523. The energy differences between the ground and the isomeric states are found to be 0.48 and 1.84 
MeV for the nearest consucative minimas. For SHF, the minima appear at around $\beta_2$= -0.459,-0.159 and 0.511. The 
intrinsic excited state energy differences are 1.30 and 0.48 MeV. From the figure it is clear that the mimima and the maxima 
in both the RMF and SHF are qualitatively similar. The absolute value differ by a constant factor from one another, i.e., 
if we scale the lower curve by, say, a scaling factor c= 1.0075 then both the curves will coincide with each other. This 
difference in energy is also reflected in the binding energy calculations of this nucleus in an isotopic chain, which will 
be discussed in the following subsection.  

\begin{table*}
\caption{The SHF(SkI4) and the RMF(NL3) results for binding energy BE, two-neutron separation energy $S_{2n}$ and
the quadrupole deformation parameter $\beta_{2}$, compared with the Finite Range Droplet Model (FRDM) data \cite{moll97}.
The energy is in MeV.
}
\begin{tabular}{|c|c|c|c|c|c|c|c|c|c|c|}
\hline
&\multicolumn{3}{c|}{SHF(SkI4 parameter set)}&\multicolumn{3}{c|}{RMF(NL3 parameter set)}&\multicolumn{3}{c|}{FRDM results}\\
\hline
Nucleus& BE  & $S_{2n}$ & $\beta_{2}$ & BE  & $S_{2n}$ & $\beta_{2}$ & BE & $S_{2n}$ & $\beta_{2}$ \\
\hline
294 & 2062.49 & 16.29 & 0.534 & 2045.52 & 16.71 & 0.530 & 2053.16 &  &  -0.155\\
296 & 2078.46 & 15.94 & 0.529 & 2061.74 & 16.21 & 0.527 & 2068.99 & 15.84 & -0.130\\
298 & 2093.81 & 15.34 & 0.526 & 2077.44 & 15.70 & 0.536 & 2084.26 & 15.26 & -0.096\\
300 & 2108.67 & 14.81 & 0.526 & 2092.62 & 15.18 & 0.548 & 2099.64 & 15.38 & 0.009  \\
302 & 2123.01 & 14.34 & 0.529 & 2107.30 & 14.68 & 0.562 & 2113.98 & 14.34 & 0.418\\
304 & 2136.83 & 13.82 & 0.545 & 2121.47 & 14.17 & 0.603 & 2126.87 & 12.89 & 0.000\\
306 & 2150.03 & 13.20 & 0.556 & 2135.23 & 13.76 & 0.608 & 2139.43 & 12.56 & 0.000\\
308 & 2162.49 & 12.45 & 0.560 & 2148.30 & 13.08 & 0.618 & 2150.84 & 11.41 & 0.001\\
310 & 2174.49 & 12.00 & 0.571 & 2160.66 & 12.35 & 0.641 & 2162.05 & 11.22 & 0.003\\
312 & 2187.10 & 12.62 & 0.584 & 2172.58 & 11.92 & 0.742 & 2173.42 & 11.36 & 0.005\\
314 & 2199.12 & 12.02 & 0.594 & 2184.17 & 11.59 & 0.739 & 2184.67 & 11.25 & 0.006\\
316 & 2210.49 & 11.37 & 0.595 & 2195.39 & 11.22 & 0.736 & 2195.74 & 11.07 & 0.007\\
318 & 2221.02 & 10.65 & 0.588 & 2206.30 & 10.91 & 0.722 & 2214.11 & 18.37 & 0.541\\
320 & 2231.23 & 10.21 & 0.575 & 2216.96 & 10.67 & 0.728 & 2224.88 & 10.76 & 0.543\\
\hline
\end{tabular}
\label{Table 1}
\end{table*}

\subsection{Binding energy and Two-neutron separation energy }
 
Fig. 2 shows the calculated binding energy, obtained in both the SHF and RMF formalisms. We notice that, similar to the PES, 
the binding energy obtained in the RMF model also over-estimates the SHF result by a constant factor. In other words, here 
also the multiplication by a constant factor 'c' will make the two curves overlap with one another. This means that a slight
modification of the parameter set of one formalism can predict the binding energy similar to that of the other. 

Table I shows a comparison of the calculated binding energies with the Finite Range Droplet Model (FRDM) predictions of 
Ref. \cite{moll97}, wherever possible. The two-neutron separation energy $S_{2n}$(N,Z)=BE(N,Z)-BE(N-2,Z) is also listed in
Table I. From the table, we find that the microscopic binding energies and the $S_{2n}$ values agree well with the 
macro-microscopic FRDM calculations. 

The comparison of $S_{2n}$ for the SHF and RMF with the FRDM result are further shown in Fig. 3, which shows clearly that 
the two $S_{2n}$ values coincide remarkably well, except at mass A=318 which seems spurious due to some error somewhere in 
the case of FRDM. Apparently, the $S_{2n}$ decrease gradually with increase of neutron number, except for the noticeable 
kinks at A=294 (N=172) and 312 (N=190) in RMF, and at A=304 (N=182) and 308 (N=186) in FRDM.  Interestingly, these neutron 
numbers are close to either N=172 or 184 magic numbers. However, the SHF results are smooth.

\begin{figure}[ht]
\vspace{1.15cm}
\begin{center}
\includegraphics[width=1.0\columnwidth]{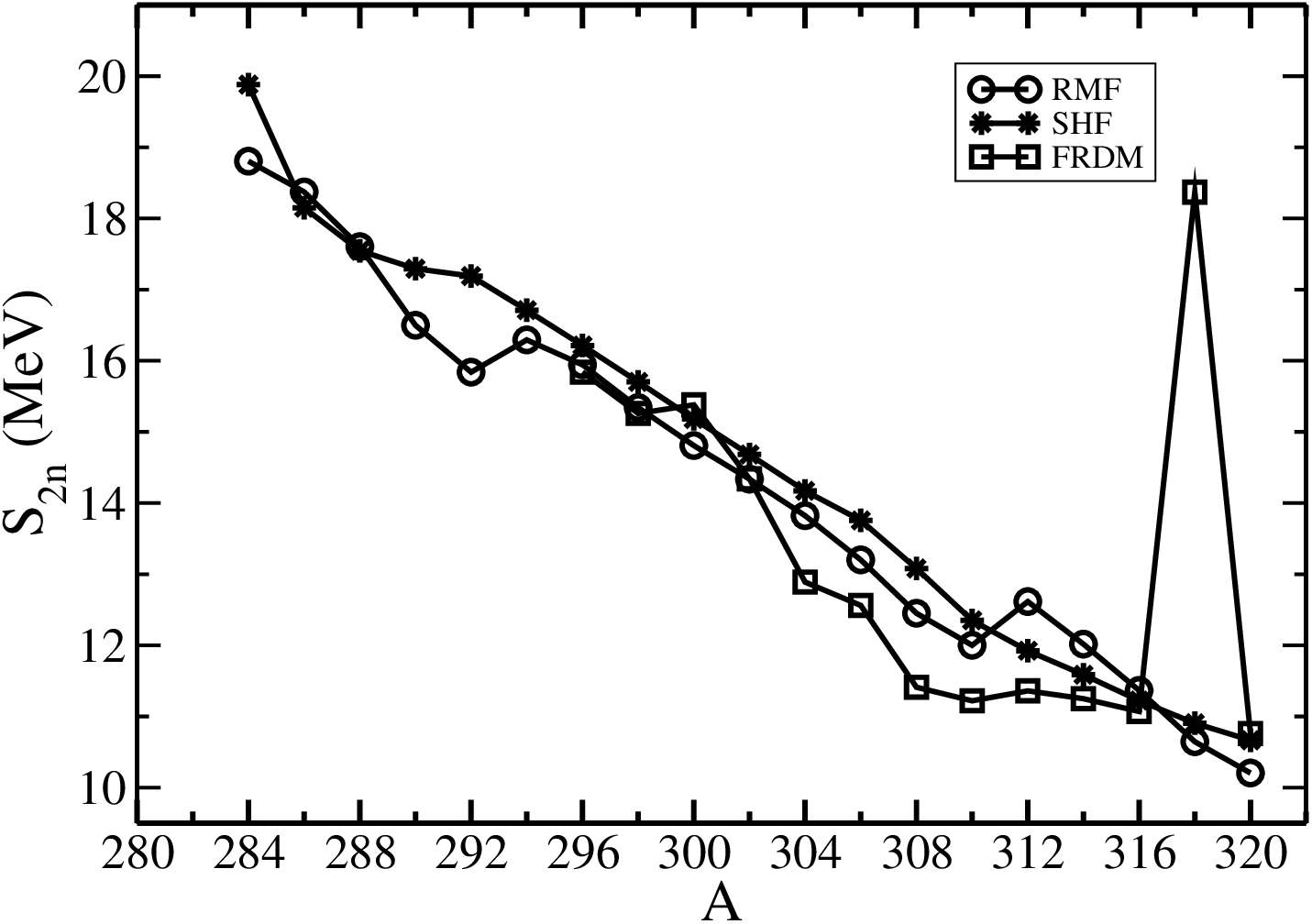}
\caption{The two-neutron separation energy ${S_2n}$ for $^{282-320}$122 nuclei, obtained from SHF(SkI4) and RMF(NL3) 
formalisms, and compared with the FRDM results \cite{moll97}, whereever available.
}
\end{center}
\label{Fig. 3}
\end{figure}

The binding energy per particle for the isotopic chain is also plotted in Fig. 4. We notice that here again the SHF and RMF 
curves could be overlapped with one another through a constant scaling factor, and the FRDM calculation lie in between 
these two calculations. This means, qualitatively, all the three curves show a similar behavior. However, unlike the BE/A 
curve for SHF or RMF, the FRDM results do not show the regular behaviour. In general, the BE/A start increasing with the 
increase of mass number A, reaching a peak value at A=302 for all the three formalisms. This means that $^{302}$122 is the 
most stable element from th binding energy point of view. Interestingly, $^{302}$122 is situated towards the neutron 
deficient side of the isotopic series of Z=122, and could be taken as a suggestion to synthesize this superheavy nucleus
experimently.

\begin{figure}[ht]
\vspace{1.0cm}  
 \begin{center}
\includegraphics[width=1.0\columnwidth]{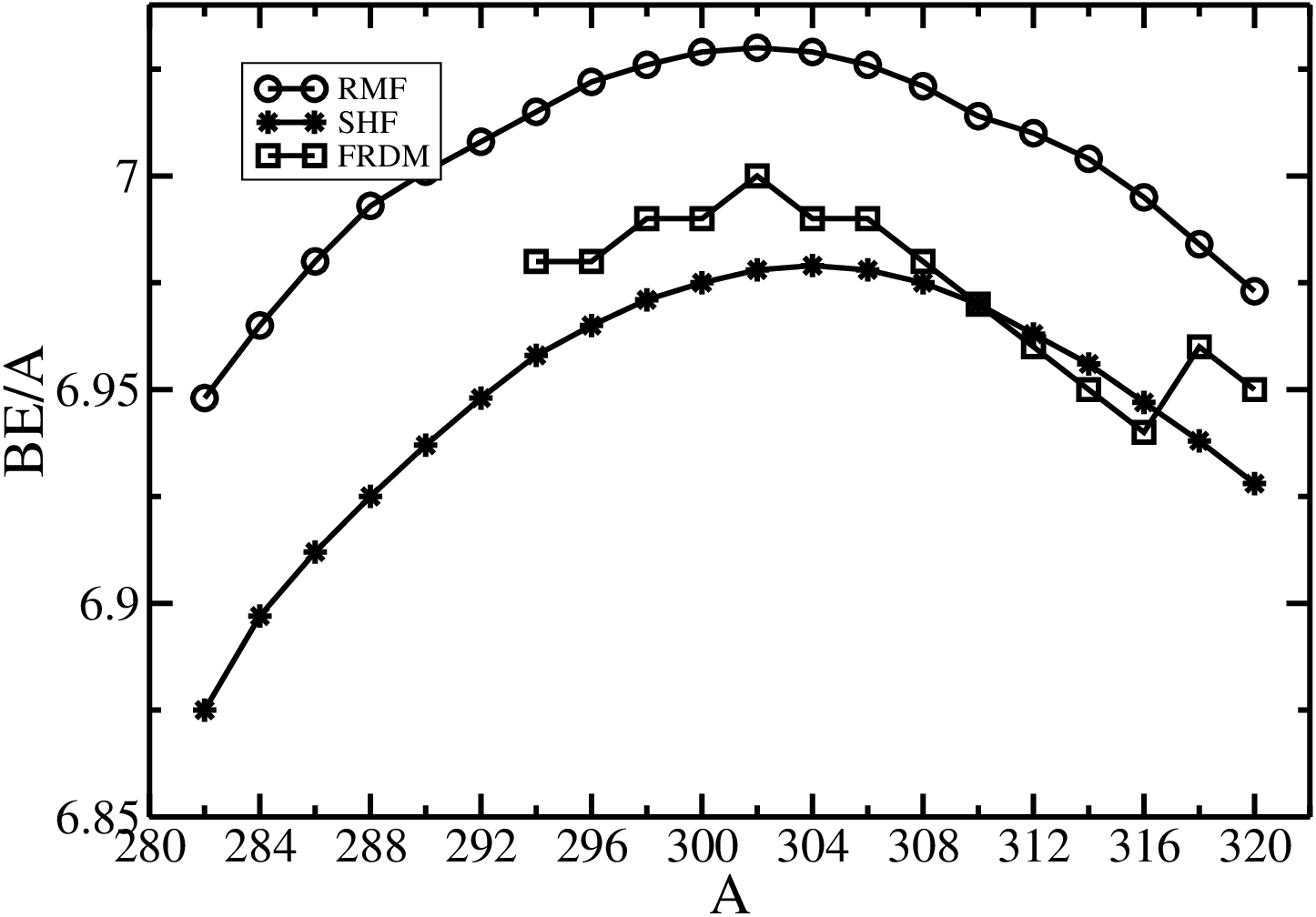}
\caption{The binding energy per particle BE/A for the superheavy isotopes $^{282-320}$122, obtained in SHF(SkI4) and 
RMF(NL3) formalisms, compared with the FRDM results \cite{moll97}, whereever available.}                                                         
\end{center}
\label{Fig. 4}
\end{figure}

\begin{figure}[ht]
\vspace{0.30cm}
\begin{center}
\includegraphics[width=1.0\columnwidth]{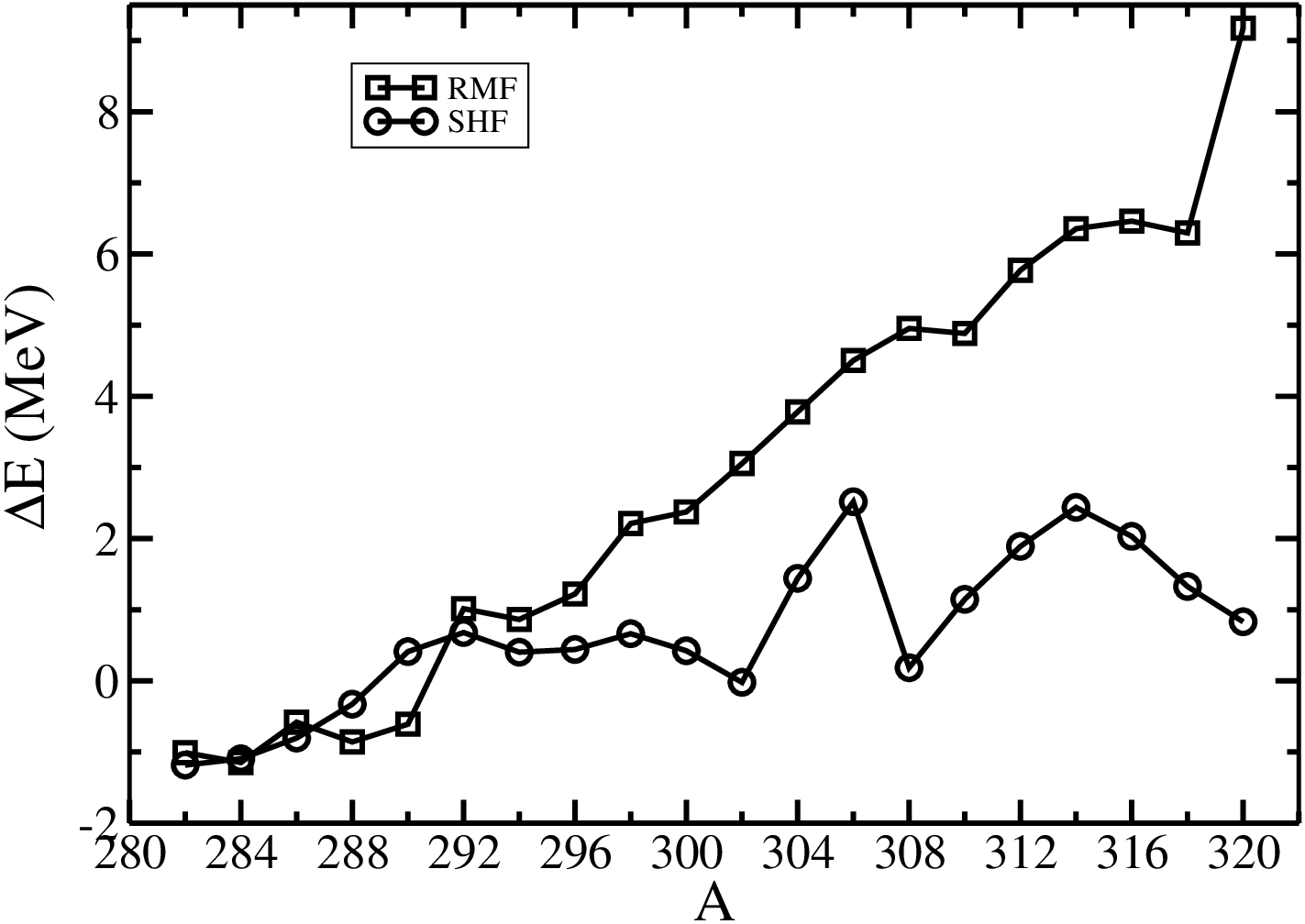}
\caption{The energy difference between the ground state and the first excited state both in nonrelativistic SHF(SkI4) and 
relativistic RMF(NL3) formalisms.
}
\end{center}
\label{Fig. 5}
\end{figure}
  
Also, we have calculated the "free solutions" for the whole isotopic chain, both in prolate and oblate deformed 
configurations. In many cases, we find low lying excited states. As a measure of the energy difference between the ground
band and the first excited state, we have plotted in Fig. 5 the binding energy difference $\triangle E$ between the two
solutions, noting that the maximum binding energy solution refers to the ground state and all other solutions to the 
intrinsic excited state(s). From Fig. 5, we notice that in RMF calculations, the energy difference $\triangle E$ is small 
for neutron-deficient isotopes, but it increases with the increase of mass number A in the isotopic series. On the other 
hand, in SHF formalism, $\triangle E$ value remains small throughout the isotopic chain. This later result means to suggest 
that the ground state can be changed to the excited state and vice-versa by a small change in the input, like the pairing 
strength, etc., in the calculations. In any case, such a phenomenon is known to exist in many other regions of the periodic 
table.

\begin{figure}[ht]
\vspace{0.75cm}
\begin{center}
\includegraphics[width=1.0\columnwidth]{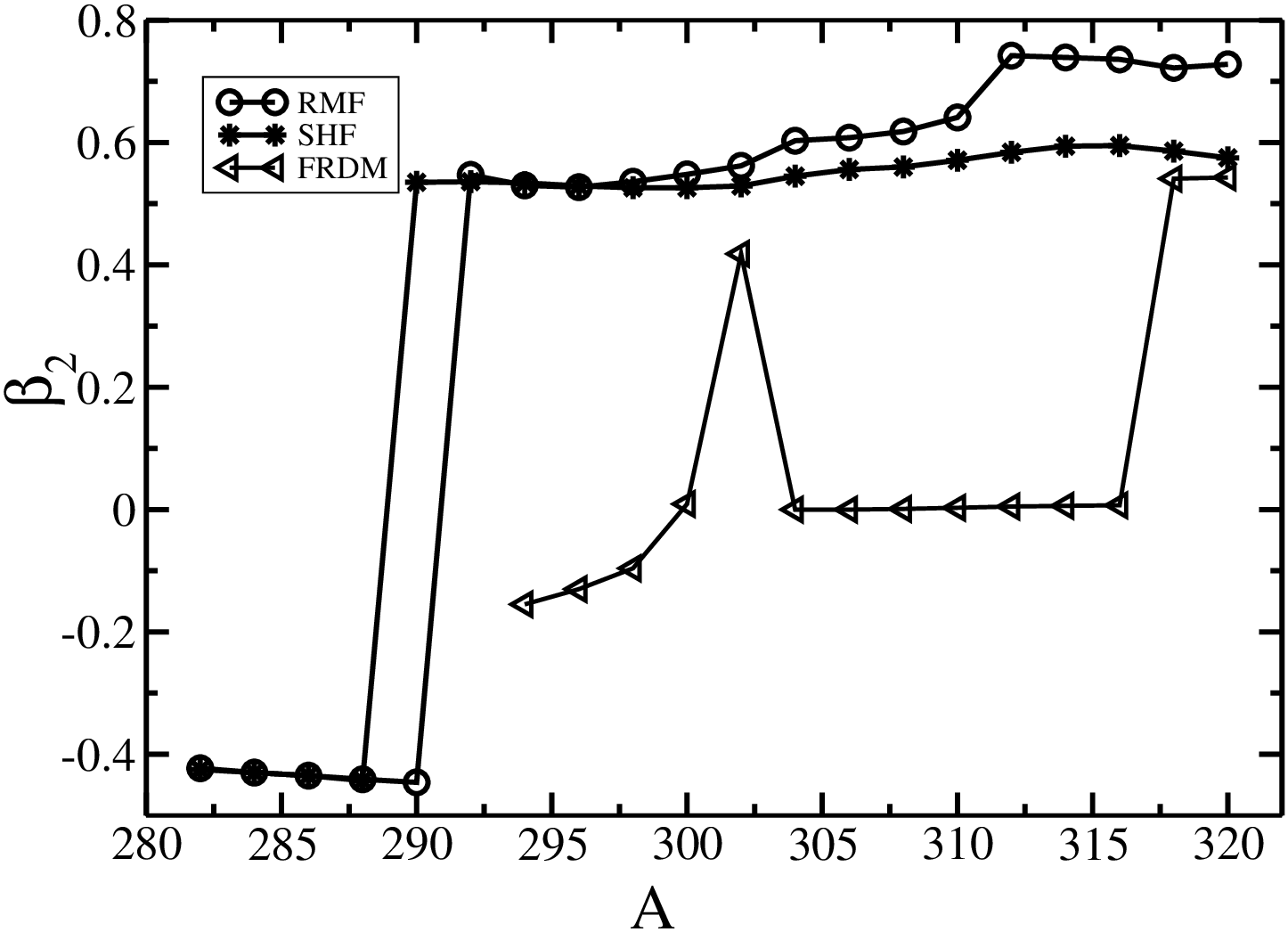}
\caption{Comparision of quadrupole deformation parameter obtained from nonrelativistic SHF(SkI4) and relativistic mean 
field formalism RMF(NL3), compared with the FRDM results \cite{moll97}, whereever available. }
\end{center}
\label{Fig. 6}
\end{figure}

\subsection{Quadrupole deformation parameter}

The quadrupole deformation parameter $\beta_2$, for both the ground and first excited states, are also determined within 
the two formalisms. In some of the earlier RMF and SHF calculations, it was shown that the quadrupole moment obtained 
from these theories reproduce the experimental data pretty well 
\cite{patra1,cha97,rei95,sero86,ring90,lala97,patra2,cha98,brown98}. We have seen in Fig. 1 that both the ground-state and 
intrinsic excited quadrupole deformation parameters for SHF and RMF results agree well with each other (the same is true
for ``free solutions", not shown here). However, the ground-state (g.s.) quadrupole deformation parameter $\beta_2$ plotted 
in Fig. 6 for SHF and RMF, and compared with FRDM results \cite{moll97}, show that the FRDM results differ strongly. Both in 
the SHF and RMF results, we find highly deformed oblates solutions in the g.s. confuguration for isotopes near the low mass 
region. Then, with increase of mass number there is a shape change from highly oblate to highly prolate in both SHF and RMF 
models. Interestingly, most of the isotopes are superdeformed in their g.s. confugurations, and due to the shape 
co-existance proporties of these isotopes, some time it is possible that the g.s. could be the hyperdeformed 
solution.

\begin{table*}
\caption{The $Q_{\alpha}$ and $T_{\alpha}$ calculated on the SHF(SkI4) and the RMF(NL3) models, and compared with the 
Finite Range Droplet Model (FRDM) results \cite{moll97}, whereever available. The energy is in MeV.
}
\begin{tabular}{|c|c|c|c|c|c|c|c|c|c|c|}
\hline
&&\multicolumn{3}{c|}{SHF(SkI4 parameter set)}&\multicolumn{3}{c|}{RMF(NL3 parameter set)}&\multicolumn{3}{c|}{FRDM results}\\
\hline
Nucleus & Z & BE & $Q_{\alpha}$ & $T_{\alpha}$ & BE & $Q_{\alpha}$ & $T_{\alpha}$ & BE & $Q_{\alpha}$ & $T_{\alpha}$\\
\hline
292 & 122 & 2028.81 & 14.31 & $10^{-7.23}$ & 2046.19 & 13.83 & $10^{-6.35}$ & &  & \\
288 & 120 & 2014.82 & 13.13 & $10^{-5.49}$ & 2031.75 & 12.35 & $10^{-3.85}$ & 2023.06 & 13.98 & $10^{-6.07}$ \\
284 & 118 & 1999.65 & 14.86 & $10^{-9.11}$ & 2015.80 & 12.87 & $10^{-5.48}$ & 2008.69 & 12.70 & $10^{-4.08}$\\
280 & 116 & 1986.21 & 13.89 & $10^{-7.93}$ & 2000.37 & 12.92 & $10^{-6.10}$ & 1993.49 & 12.42 & $10^{-5.10}$ \\
276 & 114 & 1971.80 & 12.30 & $10^{-5.37}$ & 1984.99 & 11.82 & $10^{-4.33}$ & 1977.62 & 12.33 & $10^{-5.44}$\\
272 & 112 & 1955.80 & 12.33 & $10^{-5.97}$ & 1968.51 & 11.45 & $10^{-4.07}$ & 1961.66 & 11.61 & $10^{-4.45}$\\
268 & 110 & 1939.83 & 11.86 & $10^{-5.54}$ & 1951.66 & 10.92 & $10^{-3.41}$ & 1944.97 & 10.94 & $10^{-3.47}$\\
264 & 108 & 1923.39 & 10.25 & $10^{-2.34}$ & 1934.28 & 10.19 & $10^{-2.19}$ & 1927.62 & 10.57 & $10^{-3.18}$\\
260 & 106 & 1905.34 & 9.59 & $10^{-1.10}$ & 1916.17 & 9.98 & $10^{-2.27}$ & 1909.90 & 9.93 & $10^{-2.15}$\\
256 & 104 & 1886.63 & 9.71 & $10^{-2.20}$ & 1897.85 & 7.53 & $10^{4.95}$ & 1891.53 & 8.75 & $10^{0.59}$\\
252 & 102 & 1868.04 & 8.71 & $10^{0.02}$ & 1877.08 & 8.02 & $10^{2.32}$ & 1871.98 & 8.35 & $10^{1.19}$\\
248 & 100 & 1848.45 & 7.34 & $10^{4.08}$ & 1856.80 & 7.18 & $10^{4.72}$ & 1852.03 & 7.64 & $10^{2.91}$\\
244 & 98 & 1827.49 & 7.37 & $10^{3.14}$ & 1835.68 & 6.85 & $10^{5.26}$ & 1831.38 & 6.90 & $10^{5.01}$\\
240 & 96 & 1806.56 & 6.63 & $10^{3.34}$ & 1814.23 & 5.91 & $10^{8.82}$ & 1809.98 & 6.52 & $10^{5.81}$\\
236 & 94 & 1784.89 & 6.10 & $10^{5.90}$ & 1791.84 & 5.64 & $10^{9.26}$ & 1788.21 & 5.77 & $10^{8.54}$\\
232 & 92 & 1762.69 & 6.09 & $10^{5.98}$ & 1768.19 & 5.54 & $10^{8.82}$ & 1754.15 & 5.14 & $10^{11.18}$\\
\hline
\end{tabular}
\end{table*}

\subsection{Nuclear radii}

The root mean square (rms) radius for proton ($ r_{p}$), neutron ($ r_{n}$) and matter distribution ($r_{m}$), both 
in SHF and RMF formalisms, is shown in Fig. 7. The upper pannel is for the SHF and the lower one for the RMF calculations. 
As expected, the neutron and matter distribution radius increases with increase of the neutron number. Although, the 
proton number Z=122 is constant in the isotopic series, the value of $r_{p}$ also increase as shown in the figure. 
This trend is similar in both the formalisms. A minute inspection of the figure shows that, in RMF calculation, the radii 
show a jump at A=312 (N=190) after the monotonous increase of radii till A=310. Note that a similar trend was observed in 
RMF calculations for $S_{2n}$ (see, Fig. 3).

\begin{figure}[ht]
\vspace{0.65cm}
\begin{center}
\includegraphics[width=1.0\columnwidth]{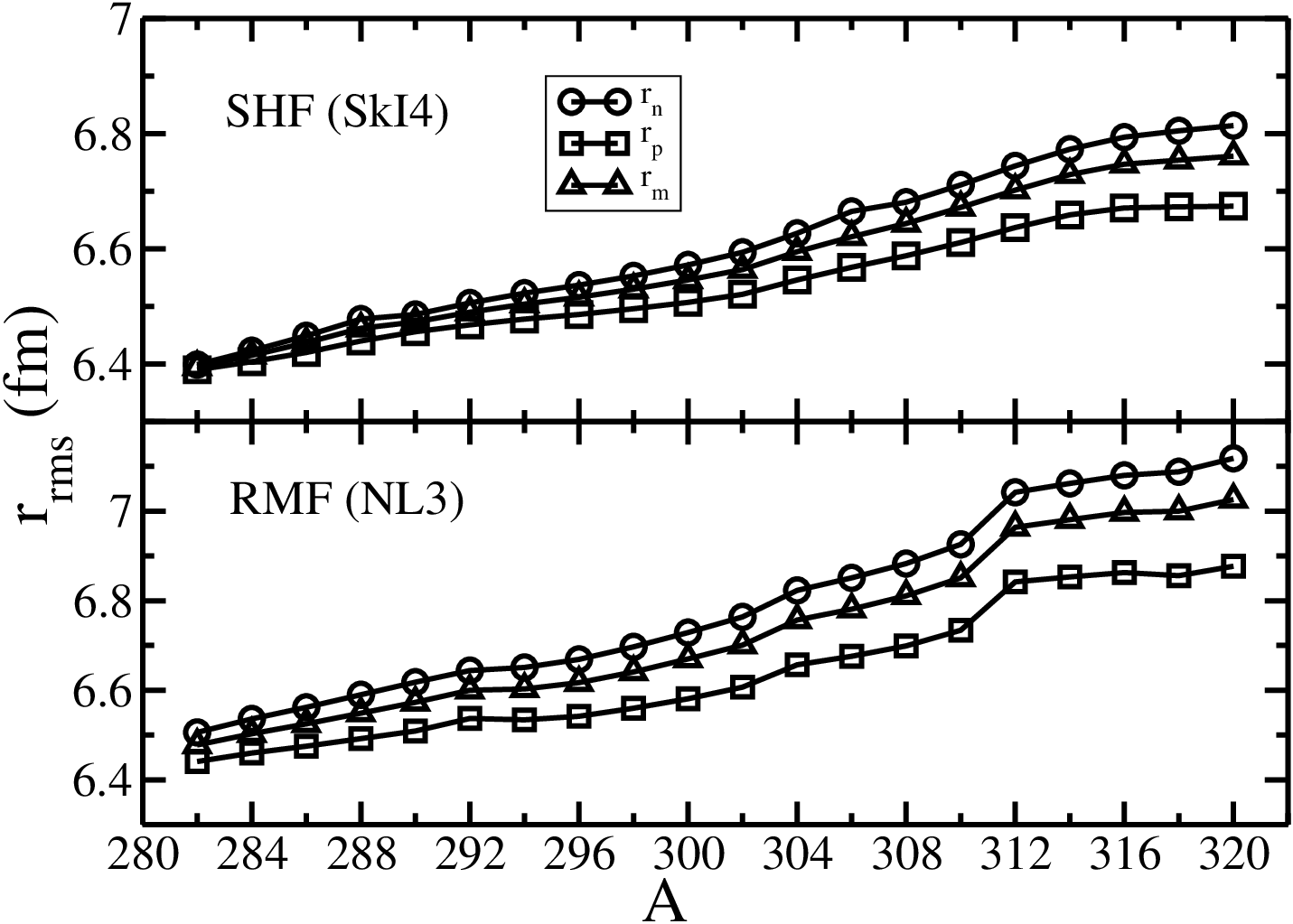}
\caption{The rms radii of proton ($r_p$), neutron ($r_n$) and matter ($r_m$) distribution for $^{282-320}$122 nuclei using 
nonrelativistic SHF(SkI4) and relativistic mean field formalism RMF(NL3).}
\end{center}
\label{Fig. 7}
\end{figure}

\begin{figure}[ht]
\vspace{0.65cm}
\begin{center}
\includegraphics[width=1.0\columnwidth]{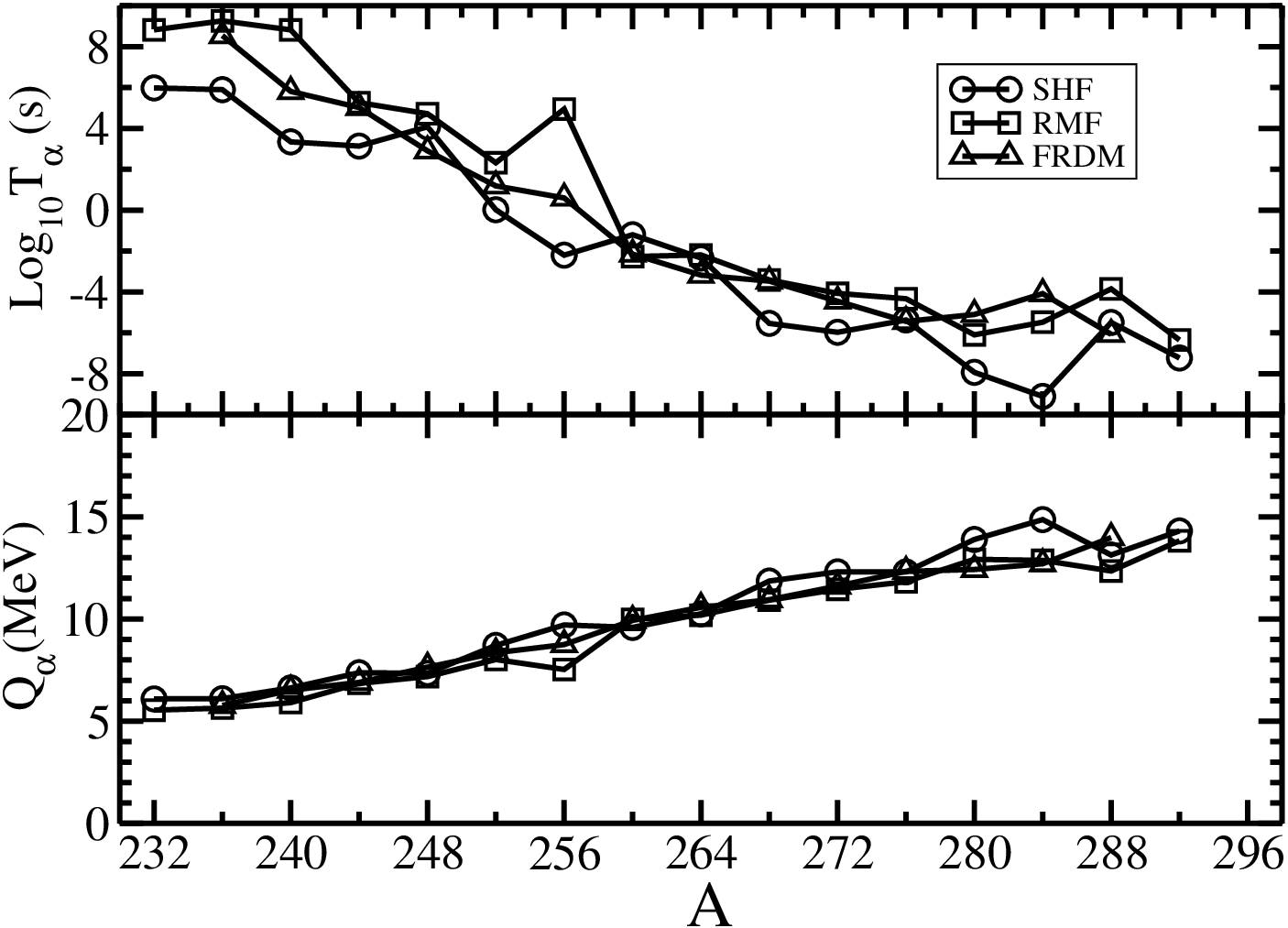}
\caption{The half-life time $T_{\alpha}$ and the $Q_{\alpha}$ energy for $^{292}$122 nucleus, using the non-relativistic
SHF(SkI4), the relativistic mean field formalism RMF(NL3), and the FRDM data \cite{moll97}. }
\end{center}
\label{Fig. 8}
\end{figure}

\subsection{The $Q_{\alpha}$ energy and the decay half-life $T_{\alpha}$}

We choose the nucleus $^{292}$122 (Z=122, N=170) for illustrating our calculations of the $\alpha$-decay chain and the 
half-life time $T_{\alpha}$. The $Q_{\alpha}$ energy is obtained from the relation \cite{patra23}:
$$Q_{\alpha}(N, Z)=BE(N, Z)-BE(N-2, Z-2)-BE(2, 2).$$
Here, $BE(N, Z$)is the binding energy of the parent nucleus with neutron number $N$ and proton number $Z$, $BE(2, 2)$ is 
the binding energy of the $\alpha$-particle ($^4He$) and $BE(N-2, Z-2)$ is the binding energy of the daughter nucleus after 
the emission of an $\alpha$-particle. 

The binding energy of the parent and daughter nuclei are obtained by using both the RMF and SHF formalisms. Our predicted 
results are compared in Table  II with the finite range droplet model (FRDM) calculation of Ref. \cite{moll97}. The 
$Q_{\alpha}$ values are then calculated, also shown in Table II and in lower panel of Fig. 8. Then, the half-life 
$Log_{10}T_{\alpha}(s)$ are estimated by using the phenomenological formulla of Viola and Seaborg \cite{viol01}:
$$Log_{10}T_{\alpha}(s)=\frac {aZ-b}{\sqrt{Q_{\alpha}}}-(cZ+d)$$
where Z is the atomic number of parent nucleus, $a$=1.66175, $b$=8.5166, $c$=0.20228 and $d$=33.9069. The calculated 
$Log_{10}T_{\alpha}(s)$ are also given in Table II and in upper panel of Fig. 8. 

From Fig. 8, we notice that the calculated values for both $Q_{\alpha}$ and $T_{\alpha}(s)$ agree quite well with the FRDM 
predictions. For example, the value of $T_{\alpha}$, in  both the FRDM and RMF coincides for the $^{264}Hs$ nucleus.  
Similarly, for $^{276}114$, the SHF prediction matches the FRDM result. Possible shell structure effects in $Q_{\alpha}$,
as well as in $T_{\alpha}(s)$, are noticed for the daughter nucleus A=256 (Z=104, N=152) and 284 (Z=118, N=166) in SHF and 
for A=256 (A=104, N=152) and 288 (Z=120, N=168) in RMF calculations. Note that some of these proton or neutron numbers refer 
to either observed or prediced magic numbers.

\section{Summary}

Concluding, we have calculated the binding energy, rms radius and quadrupole deformation parameter for the possibly 
discovered Z=122 superheavy element recently. From the calculated binding energy, we also estimated the two-neutron 
separation energy for the isotopic chain. We have employed both the SHF and RMF formalisms in order to see the formalism 
dependence of the results. We found qualitatively similar predictions in both the techniques. A shape change from oblate 
to prolate deformation is observed with increase of isotopic mass number at A=290. The ground-state structures are highly 
deformed which are comparable to superdeformed or hyperdeformed solutions, in agreement with the observations of Ref. 
\cite{marinov09} for the superheavy region. From the binding energy analysis, we found that the most stable isotope in the 
series is $^{302}$122, instead of the observed $^{292}$122, consideed to be a neutron-deficient nucleus. Our predicted 
$\alpha$-decay energy $Q_{\alpha}$ and half-life time $T_{\alpha}$ agree nicely with the FRDM calculations. Some shell 
structure is also observed in the calculated quantities at N=172 or 190 for RMF and at N=182-186 for SHF calculations for 
the various isotopes of Z=122 nucleus.

\section*{Acknowledgments}
This work is supported in part by Council of Scientific \& Industrial Research (Project No.03(1060)06/EMR-II, and by the 
Department of Science and Technology (DST), Govt. of India (Project No. SR/S2/HEP-16/2005).

\end{document}